\documentclass[onecollarge]{svjour2}       
\smartqed  
\usepackage{graphicx}
\usepackage{epsfig}
\usepackage{amsmath,amssymb}
 \newcommand{\be}{\begin{eqnarray}}
 \newcommand{\ee}{\end{eqnarray}}
 \newcommand{\al}{\alpha}
 \newcommand{\ga}{\gamma}
 \newcommand{\de}{\delta}
 \newcommand{\De}{\Delta}
 \newcommand{\La}{\Lambda}
 \newcommand{\Om}{\Omega}
 \newcommand{\dd}{\partial}

\journalname{General Relativity and Gravitation}
\begin{document}
\title{Dark Energy and Dark Gravity: Theory Overview }

\titlerunning{Dark Energy and Dark Gravity }
\author{Ruth Durrer \and Roy Maartens}
\authorrunning{R. Durrer \& R. Maartens}
\institute{R. Durrer \at D\'epartment de Physique Th\'eorique,
              Universit\'e de Gen\`eve, 24 Quai E. Ansermet, 1211
              Gen\`eve 4, Switzerland\\
              \email{ruth.durrer@physics.unige.ch} \and
           R. Maartens \at Institute of Cosmology \& Gravitation,
              University of Portsmouth, Portsmouth PO1 2EG, UK\\
              \email{roy.maartens@port.ac.uk}
}

\maketitle

\begin{abstract}
Observations provide increasingly strong evidence that the
universe is accelerating. This revolutionary advance in
cosmological observations confronts theoretical cosmology with a
tremendous challenge, which it has so far failed to meet.
Explanations of cosmic acceleration within the framework of
general relativity are plagued by difficulties. General
relativistic models are nearly all based on a dark energy field
with fine-tuned, unnatural properties. There is a great variety of
models, but all share one feature in common -- an inability to
account for the gravitational properties of the vacuum energy.
Speculative ideas from string theory may hold some promise, but it
is fair to say that no convincing model has yet been proposed. An
alternative to dark energy is that gravity itself may behave
differently from general relativity on the largest scales, in such
a way as to produce acceleration. The alternative approach of
modified gravity (or dark gravity) provides a new angle on the
problem, but also faces serious difficulties, including in all
known cases severe fine-tuning and the problem of explaining why
the vacuum energy does not gravitate. The lack of an adequate
theoretical framework for the late-time acceleration of the
universe represents a deep crisis for theory -- but also an
exciting challenge for theorists. It seems likely that an entirely
new paradigm is required to resolve this crisis.
\end{abstract}

\section{INTRODUCTION}

The current ``standard model" of cosmology is the inflationary
cold dark matter model with cosmological constant, usually called
LCDM, which is based on general relativity and particle physics
(i.e., the Standard Model and its minimal supersymmetric
extensions). This model provides an excellent fit to the wealth of
high-precision observational data, on the basis of a remarkably
small number of cosmological parameters~\cite{data}. In
particular, independent data sets from cosmic microwave background
(CMB) anisotropies, galaxy surveys and supernova luminosities,
lead to a consistent set of best-fit model parameters -- which
represents a triumph for LCDM.

The standard model is remarkably successful, but we know that its
theoretical foundation, general relativity, breaks down at high
enough energies, usually taken to be at the Planck scale,
\begin{equation}
E \gtrsim M_p \sim 10^{16}\,\mbox{TeV}\,.
\end{equation}
The LCDM model can only provide limited insight into the very
early universe. Indeed, the crucial role played by inflation
belies the fact that inflation remains an effective theory without
yet a basis in fundamental theory. A quantum gravity theory will
be able to probe higher energies and earlier times, and should
provide a consistent basis for inflation, or an alternative that
replaces inflation within the standard cosmological model (for
recent work in different directions, see e.g. Refs.~\cite{qg}).

An even bigger theoretical problem than inflation is that of the
late-time acceleration in the expansion of the universe~\cite{de}.
In terms of the fundamental energy density parameters, the data
indicates that the present cosmic energy budget is given by
 \be
\label{olom} \Omega_\Lambda &\equiv& {\Lambda \over 3H_0^2}\approx
0.75\,,~~ \Omega_{m} \equiv {8\pi G\rho_{m0} \over 3H_0^2}\approx
0.25\,, ~~
\Omega_{K} \equiv {-K\over H_0^2} \approx 0 \,, \\
\Omega_r &\equiv& {8\pi G\rho_{r0} \over 3H_0^2}\approx 8\times
10^{-5}\,.
 \ee
Here $H_0$ is the present value of the Hubble parameter, $\La$ is the
cosmological constant, $K$ is spatial curvature, $\rho_{m0}$ is the
present matter density and $\rho_{r0}$ is the present radiation density.
$G$ is Newton's constant. The Friedman equation is
\begin{eqnarray}
\left({\dot a\over a}\right)^2 \equiv H^2 &=& {8\pi G\over
3}(\rho_m+\rho_r)+{\Lambda \over 3}-{K \over a^2} \nonumber\\ &=&
H_0^2\left[\Omega_m(1+z)^3+ \Omega_r(1+z)^4 +\Omega_\Lambda+
\Omega_K(1+z)^2 \right], \label{h}
\end{eqnarray}
where $a$ denotes the scale factor which is related to the
cosmological redshift by $z=a^{-1}-1$. We normalize the present scale
factor to $a_0=1$. Together with the energy conservation equation this implies
\begin{equation}
{\ddot a \over a} = -{4\pi G \over 3} \left(\rho_m+2\rho_r\right)
+ {\Lambda \over 3} \,. \label{acc}
\end{equation}
The observations, which together with Eq.~(\ref{h}) lead to the
values given in Eq.~(\ref{olom}), produce via Eq.~(\ref{acc}) the
dramatic conclusion that the universe is currently accelerating,
 \be
\ddot a_0>0\,.
 \ee

This conclusion holds only if the universe is (nearly) homogeneous
and isotropic, i.e., a Friedmann-Lema\^\i tre model. In this case
the distance to a given redshift $z$, and the time elapsed since
that redshift, are tightly related via the only free function of
this geometry, $a(t)$. If the universe instead is isotropic around
us but not homogeneous, i.e., if it resembles a
Tolman-Bondi--Lema\^\i tre solution with our galaxy cluster at the
centre, then this tight relation between distance and time for a
given redshift would be lost and present data would not
necessarily imply acceleration. This point is discussed in detail
in the contribution by Enqvist~\cite{kari}. Of course isotropy
without homogeneity violates the Copernican principle as it puts
us in the centre of the Universe. However, it has to be stressed
that up to now observations of homogeneity are very limited,
unlike isotropy, which is firmly established. Homogeneity is
usually inferred from isotropy together with the Copernican
principle. With future data, it will in principle be possible to
distinguish observationally an isotropic but inhomogeneous
universe from an isotropic and homogeneous universe (see
e.g.~\cite{Goodman}). In the following, we disregard this
possibility and assume that the Copernican principle applies.

The data also indicate that the universe is currently (nearly)
spatially flat,
 \be
|\Omega_K|\ll 1\,.
 \ee
It is common to assume that this implies $K=0$ and to use
inflation as a motivation. However, inflation does {\em not} imply
$K=0$, but only $\Omega_K\to 0$. In the late universe, the
distinction may be negligible. But in the very early universe, a
nonzero curvature can have significant effects (see
e.g.~\cite{ellmaa}). In fact, if curvature is small but
non-vanishing, neglecting it in the analysis of Supernova data can
sometimes induce surprisingly large errors, as  discussed in the
contribution by Hlozek et al.~\cite{brucechris}.

\begin{figure*}
\begin{center}
\includegraphics[height=3.25in,width=3.in]{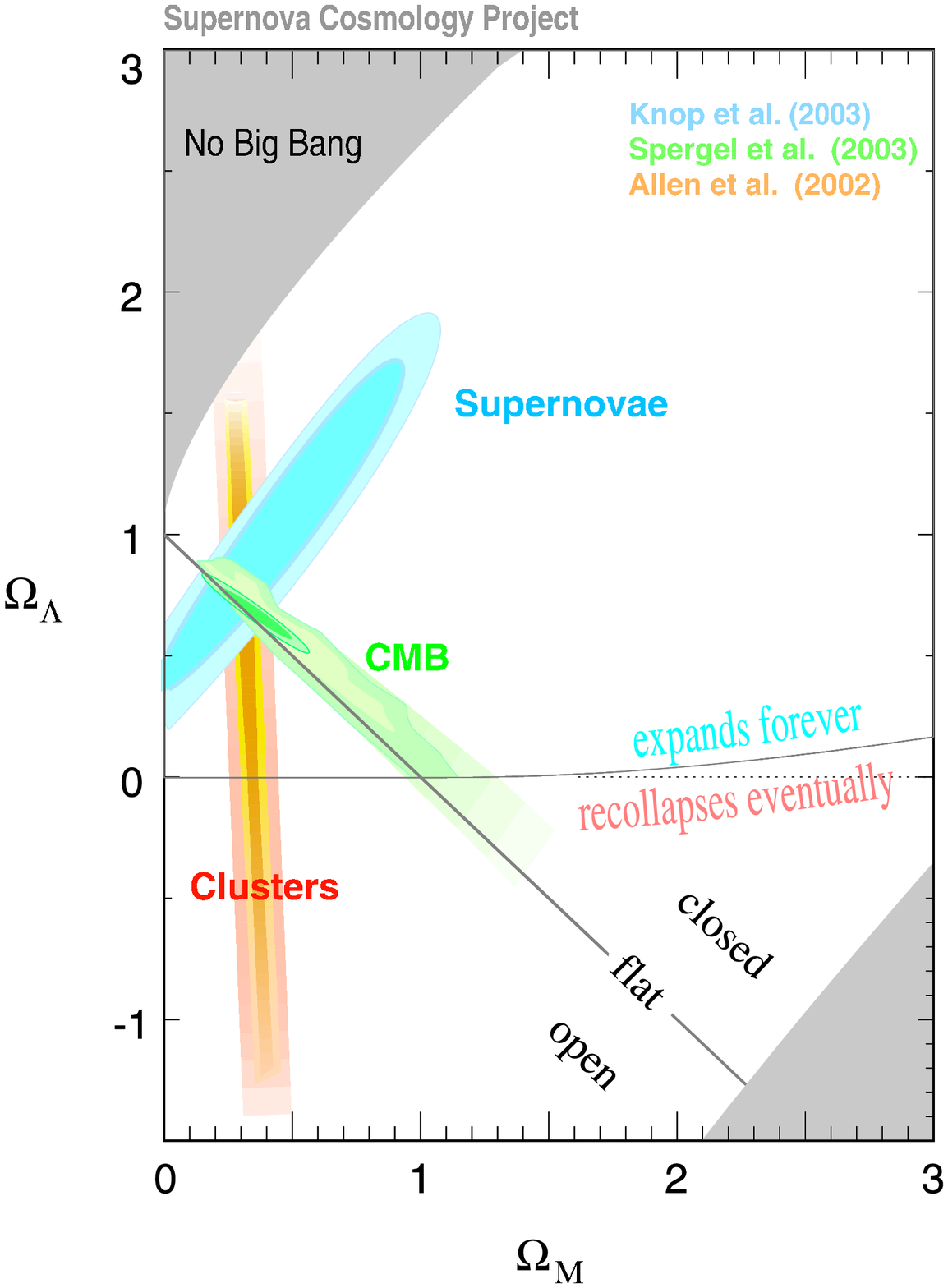}\quad
\includegraphics[height=3.25in,width=3.in]{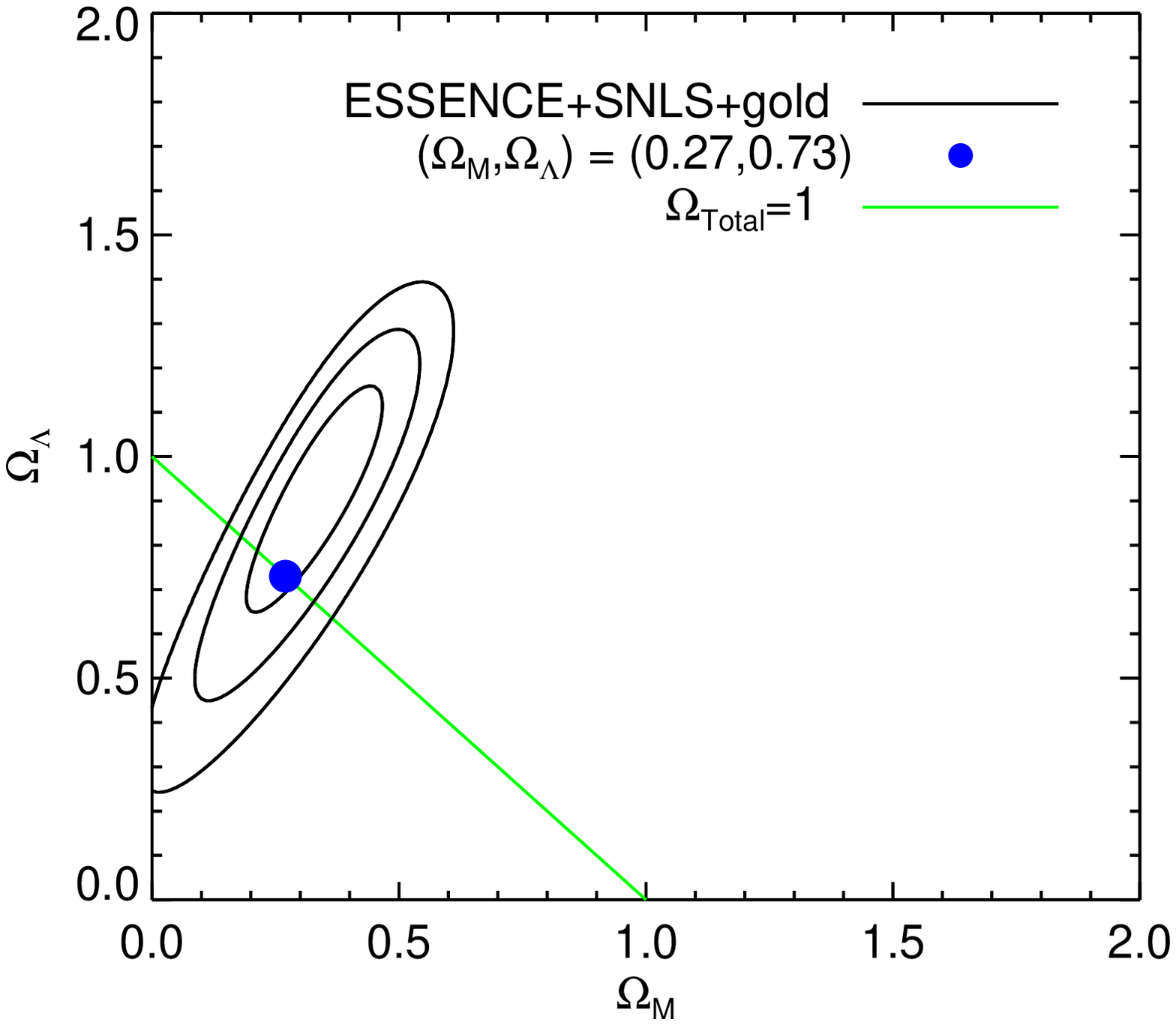}
\end{center}
\caption{Observational constraints in the
$(\Omega_{m},\Omega_\Lambda)$ plane: joint constraints (left)
(from~\cite{Knop:2003iy}); recent compilation of supernova
constraints (right) (from~\cite{Wood-Vasey:2007jb}). The line
$\Om_K=\Om_m+\Om_\La-1=0$ is also indicated.} \label{sn}
\end{figure*}

These results are illustrated in Fig.~\ref{sn} (taken
from~\cite{Knop:2003iy,Wood-Vasey:2007jb}). A detailed discussion
of the experimental aspects of the late-time acceleration is given
in the contributions by Leib\-und\-gut~\cite{Leibu},
Nichol~\cite{Nichol} and Sarkar~\cite{Sarkar}.

The simplest option is probably a cosmological constant, i.e., the
LCDM model. Even though the cosmological constant can be
considered as simply an additional gravitational constant (in
addition to Newton's constant), a cosmological constant enters the
Einstein equations in exactly the same way as a contribution from
the vacuum energy, i.e., via a Lorentz-invariant energy-momentum
tensor $T^{\rm vac}_{\mu\nu}= -(\Lambda/8\pi G) g_{\mu\nu}$. The
only observable signature of both a cosmological constant and
vacuum energy is their effect on spacetime -- and so a vacuum
energy and a classical cosmological constant cannot be
distinguished by observation. Therefore the `classical' notion of
the cosmological constant is effectively physically
indistinguishable from a quantum vacuum energy.

Even though the absolute value of vacuum energy cannot be
calculated within quantum field theory, {\em changes} in the
vacuum energy (e.g. during a phase transition) can be calculated,
and they do have a physical effect -- for example, on the energy
levels of atoms (Lamb shift), which is well known and well
measured. Furthermore, differences of vacuum energy in different
locations, e.g., between or on one side of two large metallic
plates, have been calculated and their effect, the Casimir force,
is well measured~\cite{casi}. Hence, there is no doubt about the
reality of vacuum energy. For a field theory with cutoff energy
scale $E$, the vacuum energy density scales with the cutoff as
$\rho_{\rm vac} \sim E^4$, corresponding to a cosmological
constant $\La_{\rm vac} =\rho_{\rm vac}/(8\pi G)$.  If $E=M_{p}$,
this yields a renormalization of the'cosmological constant' of about $\La_{\rm
vac} \sim 10^{38}$GeV$^2$, whereas the measured effective
cosmological constant is the sum of the 'bare' cosmological constant and
the contribution from renormalization,
 \be
\La_\mathrm{eff}= \La_{\rm vac}+\La \simeq
10^{-83}\,\mathrm{GeV}^2~. \label{v1}
 \ee
Hence a cancellation of about 120 orders of magnitude is required.
This is called the {\em fine tuning} or {\em size } problem of
dark energy: a cancellation is needed to arrive at a result which
is many orders of magnitude smaller than each of the
terms.\footnote{In quantum field theory we actually have to add
to the
  cut-off term $\La_{\rm vac} \simeq E_c^4/M_{pl}^2$ the unmeasureable
  'bare'
  cosmological constant. In this sense, the cosmological constant
  problem is a fine tuning
  between the unobservable 'bare' cosmological constant and the term
  coming from the cut-off scale.}
It is possible that the quantum vacuum energy is much smaller than
the Planck scale. But even if we set it to the lowest possible
SUSY scale, $E_{\rm susy}\sim 1$TeV, arguing that at higher
energies vacuum energy exactly cancels due to supersymmetry, the
required cancellation is still about 60 orders of magnitude. These
issues are discussed in the contributions by
Padmanabhan~\cite{Paddy} and Bousso~\cite{rafi}.

A reasonable attitude towards this open problem is the hope that
quantum gravity will explain this cancellation. But then it is
much more likely that we shall obtain directly $\La_{\rm
vac}+\La=0$ and not $\La_{\rm vac}+\La \simeq 3\rho_m(t_0)/(8\pi
G)$. This unexpected observational result leads to a second
problem, {\em the coincidence problem}: given that
 \be
\rho_\Lambda = {\Lambda_\mathrm{eff} \over 8\pi
G}=\,\mbox{constant}\,,~\mbox{ while }~\rho_m \propto (1+z)^3\,,
 \ee
why is  $\rho_\Lambda$ of the order of the {\em present} matter
density $\rho_m(t_0)$? It was completely negligible in most of the
past and will entirely dominate in the future.

Instead of a cosmological constant, one may also introduce a
scalar field or some other contribution to the energy-momentum
tensor which has an equation of state $w<-1/3$. Such a component
is called `dark energy'. So far, no consistent model of dark
energy has been proposed which can yield a convincing or natural
explanation of either of these problems. A variety of such models
is discussed in the contribution by Linder~\cite{linder}.

Alternatively, it is possible that there is no dark energy field,
but instead the late-time acceleration is a signal of a {\em
gravitational} effect. Within the framework of general relativity,
this requires that the impact of inhomogeneities somehow acts to
produce acceleration, or the appearance of acceleration (within a
Friedman-Lema\^{i}tre interpretation). One possibility is the
Tolman-Bondi--Lema\^\i tre model discussed in this
volume~\cite{kari}. Another possibility is that the `backreaction'
of inhomogeneities on the background, treated via nonlinear
averaging, produces effective acceleration. This is discussed in
the contribution by Buchert~\cite{thomas}.

A more radical version is the `dark gravity' approach, the idea that
gravity itself is weakened on large-scales, i.e., that there is an
``infrared'' modification to general relativity that accounts for
the late-time acceleration. Specific classes of models which
modify gravity are discussed in the contributions by Capozziello
and Francaviglia~\cite{CapFran} and by Koyama~\cite{koyama}.
Schematically, we are modifying the geometric side of the field
equations,
 \be
G_{\mu\nu}+G^{\rm dark}_{\mu\nu} = 8\pi G T_{\mu\nu}\,,
\label{mod}
 \ee
rather than the matter side,
 \be
G_{\mu\nu} = 8\pi G \left(T_{\mu\nu}+ T^{\rm dark}_{\mu\nu}
\right)\,,
 \ee
as in the general relativity approach. Modified gravity represents
an intriguing possibility for resolving the theoretical crisis
posed by late-time acceleration. However, it turns out to be
extremely difficult to modify general relativity at low energies
in cosmology, without violating the low-energy solar system
constraints, or without introducing ghosts and other instabilities
into the theory. Up to now, there is no convincing alternative to
the general relativity dark energy models -- which themselves are
not convincing.

The plan of the remainder of this paper is as follows. In
Section~2 we discuss constraints which one may formulate for a
dark energy or modified gravity (dark gravity) theory from basic
theoretical requirements. In Section~3 we discuss models that
address the dark energy problem within general relativity. In
Section~4 we present modified gravity models. The ideas outlined
in Sections~3 and 4 are discussed in more detail in the specific
contributions of this issue which are devoted to them. In
Section~5 we conclude.


\section{CONSTRAINING EFFECTIVE THEORIES}
\label{s:const}

The theories of both dark matter and dark energy often have very
unusual Lagrangians that cannot be quantized in the usual way,
e.g. because they have non-standard kinetic terms. We then simply
call them `effective low energy theories' of some unspecified high
energy theory which we do not elaborate. In this section, we want
to point out a few properties which we nevertheless can require of
low energy effective theories.  We first
enumerate the properties which we can require from a good basic
physical theory at the classical and at the quantum level. We then
discuss which of these requirements are inherited by low energy
effective descriptions.

\subsection{{\bf Fundamental physical theories}}
\label{sec:fund}

Here we give a minimal list of properties which we require from a
fundamental physical theory. Of course, all the points enumerated
below are open for discussion, but at least we should be aware of
what we lose when we let go of them.

In our list we start with very basic requirements which become
more strict as we go on. Even though some theorists would be able
to live without one or several of the criteria discussed here, we
think they are all very well founded. ¨Furthermore, all known current
physical theories, including string- and M-theory, do respect
them.

\setcounter{enumi}{-1}
\begin{enumerate}

\item {\bf A physical theory allows a mathematical description}\\
This is the basic idea of theoretical physics. It may well be
wrong at some stage, but it has been a working hypothesis for all
of what we call theoretical physics. If it has limitations these
may well be called the limitations of theoretical physics itself.
\vspace{0.1cm}

\item {\bf A physical theory allows a Lagrangian formulation}\\
Fundamental physical theories have a Lagrangian formulation. This
requirement is of course much stronger than the previous one. But
it has been extremely successful in the past and was the guiding
principle for the entire development of quantum field theory and
string theory in the 20th century. If we drop it, anything goes.
We can then just say the evolution of the scale factor of the
universe obeys $a(t)=At^{1/2} +Bt^{2/3} + C\exp(t/t_0)$, call this
our physical theory and fit the four parameters $A$, $B$, $C$ and
$t_0$ from cosmological data. Of course something like this does
not deserve the name 'theory'; it is simply a fit to the data.

Nevertheless sometimes fits of this kind are taken more seriously
then they should be. Some 'varying speed of light theories'
without Lagrangian formulation leave us more or less free to
specify the evolution of the speed of light during the expansion
history of the universe. However, if we introduce a Lagrangian
formulation, we realize that most of these theories are simply
some variant of scalar tensor theories of gravity, which are of
course well defined and have been studied in great detail.

If we want to keep deep physical insights like N\"other's theorem,
which relates symmetries to conservation laws, we need to require
a Lagrangian formulation for a physical theory. A basic ingredient
of a Lagrangian physical theory is that every physical degree of
freedom has a kinetic term which consists (usually) of first order
time derivatives and may also have a 'potential term' which does
not involve derivatives. In the Lagrangian formulation of a
fundamental physical theory, we do not allow for external,
arbitrarily given functions. Every function has to be a degree of
freedom of the theory so that its evolution is determined
self-consistently via the Lagrangian equations of motion, which
are of first or second order. It is possible that the Lagrangian
contains also higher than first order derivatives, but such
theories are strongly constrained by the problem of ghosts which
we mention below, and by the fact that the corresponding equations
of motion are usually described by an unbounded Hamiltonian, i.e.
the system is unstable (Ostrogradski's
theorem~\cite{Ostro,Woody}). For example, for the gravitational
Lagrangian in 4 dimensions, this means that we may only allow for
a function depending on $R$ and its derivatives, where $R$ is the
Riemann curvature scalar.  \vspace{0.1cm}

\item {\bf Lorentz invariance}\\
We also want to require that the theory be Lorentz invariant. Note
that this requirement is much stronger than demanding simply
'covariance'. It requires that there be no 'absolute element' in
the theory apart from true constants. Lorentz covariance can
always be achieved by rewriting the equations. As an example, let
us consider a Lagrangian given in flat space by
$(\dd_t\phi)^2-(\dd_x\phi)^2$. This is clearly not Lorentz
invariant. However, we can trivially write this term in the
covariant form $\alpha^{\mu\nu}\dd_\nu\dd_\mu\phi$, by setting
$(\alpha^{\mu\nu}) = {\rm diag}(1,-1,0,0)$. Something like this
should of course not be allowed in a fundamental theory. A term of
the form $\alpha^{\mu\nu}\dd_\nu\dd_\mu\phi$ is only allowed if
$\alpha^{\mu\nu}$ is itself a dynamical field of the theory. This
is what we mean by requiring that the theory is not allowed to
contain `absolute elements', i.e. it is Lorentz invariant and not
simply covariant.  \vspace{0.1cm}

\item {\bf Ghosts }\\
Ghosts are fields whose kinetic term has the wrong sign. Such a
field, instead of slowing down when it climbs up a potential, is
speeding up. This unstable situation leads to severe problems when
we want to quantize it, and it is generally accepted that one
cannot make sense of such a theory, at least at the quantum level.
This is not surprising, since quantization usually is understood
as defining excitations above some ground state, and a theory with
a ghost has no well defined ground state. Its kinetic energy has
the wrong sign and the larger $\dot\phi^2$ is, the lower is the
energy. \vspace{0.1cm}

 \item {\bf Tachyons }\\
These are degrees of freedom that have a negative mass squared,
$m^2<0$. Using again the simple scalar field example, this means
that the second derivative of the potential about the `vacuum
value' ($\phi = 0$ with $\dd_\phi V(0)=0$) is negative,
$\dd^2_\phi V(0)< 0$. In general, this need not mean that the
theory makes no sense, but rather that $\phi=0$ is a bad choice
for expanding around, since it is a maximum rather than a minimum
of the potential and therefore an unstable equilibrium.

This means also that the theory cannot be quantized around the
classical solution $\phi=0$, but it may become a good quantum
theory by a simple shift, $\phi\rightarrow \phi -\phi_0$, where
$\phi_0$ is the minimum of the potential. If the potential of a
fundamental scalar field has no minimum but only a maximum, the
situation is more severe. Then the theory is truly unstable.
\vspace{0.1cm}

The last two problems, together with the Ostrogradski instability
that appears in theories with higher derivatives, can be
summarized in the requirement that a meaningful theory needs to
have an energy functional which is bounded from below.
\vspace{0.1cm}

\item {\bf Superluminal motion and causality}\\
A fundamental physical theory which does respect Lorentz
invariance must not allow for superluminal motions. If this
condition is not satisfied, we can construct closed curves along
which a signal can propagate, in the following way.

Consider modes of a field $\phi$ which can propagate faster than
the speed of light, with velocities $v_1>1$ and $v_2>1$. Consider
a reference frame $R$ and a frame $R'$ that is boosted with
respect to $R$ by a velocity $v$ in the direction $x$, and which
coincide at the origin, $q_0$ (see Fig.~\ref{f:closed}). We choose
$v$ such that $1/v_1<v<1$. An observer in $R$ now sends a signal from
$q_0$, whose coordinates are $(t,x)=(0,0)=(t',x')$, with signal
speed $v_1$ in the direction $x$. At time $t_1$ this signal
arrives at event $q_1$, with coordinates $(t_1,x_1)$ in the
$R$-frame, where $x_1=v_1t_1$. There it is received by an observer
who is at rest with respect to $R'$, and who returns the signal
with speed $v_2'$ in the direction $-x$ to event $q_2=(t_2,0)$
(see Fig.~\ref{f:closed}). We want to show that for an appropriate
choice of $v_2'$, the time $t_2$ becomes negative.

We denote positions and times in the boosted frame $R'$ with a
prime. We then have $x_2'-x_1'= v_2'(t_2'-t_1')$. Applying the
usual formulas for Lorentz transformations, we find that $0=x_2 =
\ga(x_2'+vt_2')$ and $t_2=\ga^{-1}t_2'$. On the other hand, we
have $x_1' = \ga(x_1-vt_1)=\ga(v_1-v)t_1$ and $t_1'=
\ga(t_1-vx_1)= \ga(1-vv_1)t_1$. Note that, since we require
$vv_1>1$, it follows that $t_1'$ is negative. A signal which is
travelling at a speed greater than $1/v$ in the frame $R$ is
moving towards the past in the frame $R'$. With respect to this
frame, the event $(t_1',x_1')$ at which the signal has reached
$x_1'$, is earlier than the event $(0,0)$ when it left the
position $0$. With respect to the frame $R$ the situation is
opposite: the signal left $0$ before it reached $x_1$, $t_1>0$.
The same happens if we now send back a signal to $0$ in $R'$ with
a velocity $|v_2'|> 1/v$. This signal will travel backwards in
time $t$ with respect to $R$, and will arrive before the time
$t_1$ when it was emitted. To achieve $\De t' = t_2'-t_1' =
\ga(\De t -v\De x) =\ga(1-vv_2)\De t >0$, and at the same time
$\De t<0 $, we need $vv_2>1$, hence $v_2>1/v$. As is evident from
Fig.~\ref{f:closed}, $v_2$, which is the inverse of the slope of
the straight line connecting $q_2$ and $q_1$, must be smaller than
$v_1$, which is the inverse of the slope from $q_0$ to $q_1$.
Hence we need $1/v_1<1/v_2<v<1$. For more details
see~\cite{closed}.

\begin{figure}[ht]
\centerline{ \epsfig{file=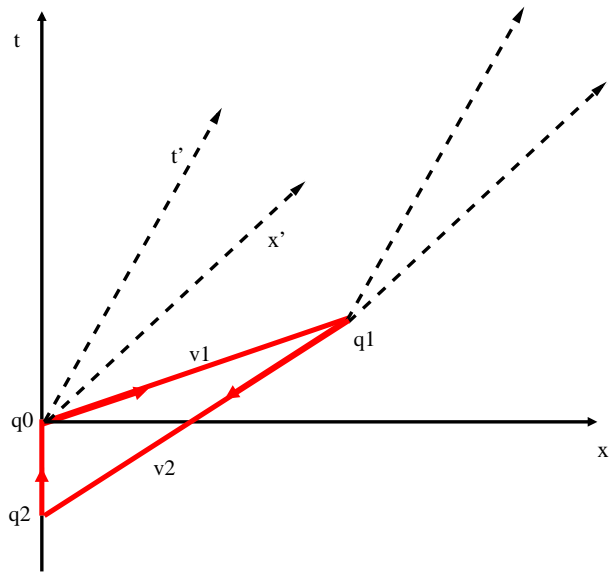, width=8cm}}
\caption{\label{f:closed} The frame R' with coordinates $(t',x')$
  moves with speed $v$ in the x-direction. A signal is sent with velocity
  $v_1$ from $q_0$ to
  $q_1$ in the frame R. Since $v_1>1/v$, this signal travels backward
  in time with respect to frame $R'$. Then a signal is sent with speed
  $v_2$ from $q_1$ to $q_2$. Since $|v_2|>1/v$, this signal, which is sent
  forward in time in frame $R'$, travels backward in time with respect
  to $R$ and can arrive at an event $q_2$ with $t_2<0$.}
\end{figure}

The loop generated in this way is not `causal' since both the
trajectory from $q_0$ to $q_1$ and the one from $q_1$ to $q_2$ are
spacelike. So we cannot speak of the formation of closed causal
loops, but it is nevertheless a closed loop along which a signal
can propagate and which therefore enables the construction of a
time machine, leading to the usual problems with causality and
entropy.

It is well known that in relativity events with spacelike
separation, like $q_0$ and $q_1$ or $q_1$ and $q_2$, have no well
defined chronology. Depending on the reference frame, one of them
is earlier than the other. Therefore superluminal motion leads to
the possibility of time machines. Hence superluminal motion is not
compatible with the equivalence of all inertial frames. Once we
allow for superluminal motion, but still require that signals can
only be sent forward in time, an event, like $q_2$ lying in the
past of $q_1$ in frame $R$ can be reached with a signal emitted in
frame $R'$, but it cannot be reached if the signal is emitted from
a source in $R$.

In a reference frame which moves with $v=1/v_1$ with respect to
$R$, a mode which propagates with velocity $v_1$ in R, has
infinite velocity. This means that the propagation equation for
this mode is no longer hyperbolic, but is elliptic, i.e. it has
become a constraint equation. In this frame the evolution of the
mode in the forward light cone of a small patch can no longer be
determined by knowing the field values (and their first
derivatives) in the small patch; the mode equation is non-local.
In all reference frames moving with a velocity $v>1/v_1$ with
respect to $R$, there exist two directions in which modes with
propagation velocity $v_1$ obey elliptic equations of motion.
Hence the Cauchy problem is not well posed in these frames. This
nullifies the equivalence of all reference frames.

At first sight one might think that a Lorentz invariant Lagrangian
will automatically forbid superluminal motions. But the situation
is not so simple. Already in the 1960s Velo and
Zwanziger~\cite{velo} discovered that generic Lorentz invariant
higher spin theories, $s\ge 1$, lead to superluminal motion. While
the equations are manifestly Lorentz invariant, their
characteristics in general do not coincide with the light cone and
can very well be spacelike. There are exceptions to this rule,
among which are Yang Mills theories for spin 1 and the linearized
Einstein equations for spin 2.

One may object to this restriction, on the grounds that general
relativity, which is certainly a theory that is acceptable (at
least at the classical level), can lead to closed causal curves,
even though it does not admit superluminal motion. Several
solutions of general relativity with closed causal curves have
been constructed in the past, see e.g
Refs.~\cite{Thorne,Gott,Ori}. But these constructions usually need
infinite energy, as in Ref.~\cite{Gott}, with two infinitely long
straight cosmic strings, or they have to violate the dominant
energy condition, as in Ref.~\cite{Thorne}, where wormholes are
used, or the closed causal curve is hidden behind an event
horizon, as in Ref.~\cite{Ori}, where an ordinary closed circle in
space is converted into a causal curve by moving it behind a
horizon which is such that the corresponding angular coordinate
becomes timelike. Nevertheless, the possibility of closed causal
curves in general relativity under certain conditions does remain
a worry, see, e.g.~\cite{Bonnor}.\vspace{0.1cm}

The situation is somewhat different if superluminal motion is only
possible in a background which breaks Lorentz-invariance. Then one
has in principle a preferred frame and one can specify that
perturbations should always propagate with the Green's function
that corresponds to the retarded Green's function in this
frame~\cite{muki}. Nevertheless, one has to accept that there will
be boosted frames relative to which the Cauchy problem for the
superluminal modes is not well defined. The physics experienced by
an observer in such a frame is most unusual (to say the least).

Causality of a theory is intimately related to  the analyticity
properties of the $S$-matrix of scattering, without which
perturbative quantum theory does not make sense. Furthermore, we
require the $S$ matrix to be unitary. Important consequences of
these basic requirements are the Kramers Kronig dispersion
relations, which are a result of the analyticity properties and
hence of causality, and the optical theorem, which is a result of
unitarity. The analyticity properties have many further important
consequences, such as the Froissart bound, which implies that the
total cross section converges at high energy~\cite{KKetal}.

\end{enumerate}

\subsection{{\bf Low energy effective theories}}
\label{sec:low}

The concept of low energy effective theories is extremely useful
in physics. As one of the most prominent examples, consider
superconductivity. It would be impossible to describe this
phenomenon by using full quantum electrodynamics with a typical
energy scale of MeV, where the energy scale of superconductivity
is milli-eV and less. However, many aspects of superconductivity
can be successfully described with the Ginzburg-Landau  theory of
a complex scalar field. Microscopically, this scalar field is to
be identified with a Cooper pair of two electrons, but this is
irrelevant for many aspects of superconductivity.

Another example is  weak interaction and four-Fermi theory.
The latter is a good approximation to weak interactions at energy
scales far below the $Z$-boson mass. Most physicists also regard
the standard model of particle physics as a low energy effective
theory which is valid below some high energy scale beyond which
new degrees of freedom become relevant, be this supersymmetry, GUT
or string theory.

\begin{figure}[ht]
\centerline{\epsfig{file=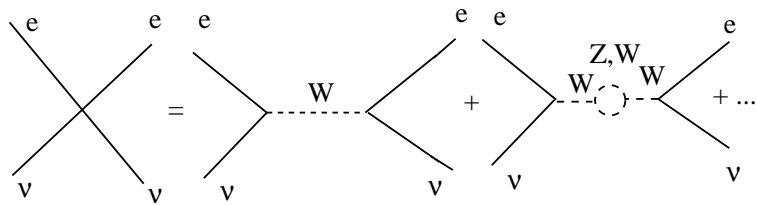, width=10cm}}
\caption{\label{f:4Fermi} The 4 Fermi interaction is the sum over
all `virtual' W and $Z$ boson exchanges. At low energy mainly the
`tree' graph contributes.}
\end{figure}

We now want to investigate which of the properties in the previous
subsection may be lost if we `integrate out' high energy
excitations and consider only processes which take place at
energies below some cutoff scale $E_c$. We cannot completely
ignore all particles with masses above $E_c$, since in the low
energy quantum theory they can still be produced `virtually',
i.e., for a time shorter than $1/E_c$. This is not relevant for
the initial and final states of a scattering process, but plays a
role in the interaction. As an example we consider 4-Fermi theory.
The vertex in the 4 fermion interaction is obtained by integration
over $W$ and $Z$ exchanges, shown in Fig.~\ref{f:4Fermi}. Even
though the final states of this theory contain only electrons and
neutrinos, the virtual presence of massive $W$'s and $Z$'s is
vital for the interaction between them.

Coming back to our list in the previous subsection, we certainly
want to keep the first point -- a mathematical description. But
the Lagrangian formulation will also survive if we proceed in a
consistent way by simply integrating out the high energy degrees
of freedom.

What about higher order derivatives in the Lagrangian? To address
this question let us briefly repeat the basic argument of
Ostrogradski's theorem simply for a (one dimensional) point
particle with time dependent position $q(t)$. If the Lagrangian
depends only on $q$ and $\dot q$, the requirement $\de S=0$
results in the ordinary Euler Lagrange equation,
 \be
\frac{d}{dt}\frac{\dd L}{\dd\dot q} -\frac{\dd L}{\dd q} = 0~.
 \ee
We can now introduce the canonical coordinates $q$ and $p\equiv
\frac{\dd L}{\dd \dot q}$. The Hamiltonian is then given by the
Legendre transform of $L$ in the variable $\dot q$,
 \be
H(q,p) = p\dot q -L\,,
 \ee
and the Euler-Lagrange equation implies the canonical equations
 \be
\dot q =\frac{\dd H}{\dd p}~, \quad \dot p =-\frac{\dd H}{\dd q}~.
 \ee
This procedure is well defined if the Lagrangian is
non-degenerate, i.e. if the equation $p\equiv \frac{\dd L}{\dd
\dot q}$ can be solved for $\dot q(q,p)$. Locally, this is
equivalent to $\frac{\dd^2 L}{\dd\dot q^2}\neq 0$. We assume the
system to be autonomous (no external time dependence). Then $H=E$
is an integral of motion, the energy of a solution, and the system
is called stable if $H$ is bounded from below. If $H$ is not
bounded from below, interactions of the system with e.g. radiation
will lead to an enormous production of radiation (massless
particles) by driving the system to lower and lower energy.

If $L$ depends also on $\ddot q$, i.e., $L(q,\dot q,\ddot q)$, the
variational principle yields
\begin{equation}
 \frac{d^2}{dt^2}\frac{\dd L}{\dd\ddot q} - \frac{d}{dt}\frac{\dd
 L}{\dd\dot q} + \frac{\dd L}{\dd q} = 0~.
\end{equation}
This is a fourth order differential equation and its solutions
depend on four initial data, $q(0), ~\dot q(0),~\ddot q(0)$ and
$\dddot q(0)$. A Hamiltonian formulation will now require four
canonical variables, which can be chosen as
\begin{equation}
q_1\equiv q~,\quad q_2 \equiv \dot q~, \quad \mbox{ and } \qquad
p_1 \equiv \frac{\dd L}{\dd\dot q}-  \frac{d}{dt}\frac{\dd
 L}{\dd\ddot q}~,\quad p_2 \equiv  \frac{\dd L}{\dd\ddot q}~.
\end{equation}
The Hamiltonian obtained by Legendre transforming the Lagrangian
with respect to the coordinates $\dot q\equiv q_2$ and $\ddot q$
yields
\begin{equation}
H(q_1,q_2,p_1,p_2) = p_1q_2+p_2\ddot q(q_1,q_2,p_2)- L\left(q_1,
q_2, \ddot q(q_1,q_2,p_2)\right) ~.
\end{equation}
This procedure is well defined if the Lagrangian is non-degenerate
in the sense that  $p_2 \equiv  {\dd L}/{\dd\ddot q}$ can be
inverted to determine $\ddot q$. Locally this requires ${\dd^2
L}/{\dd\ddot q^2} \neq 0$.

It is easy to check that the canonical equations are satisfied and
$H$ is an integral of motion. But since the Lagrangian is only a
function of three and not four variables, $p_1$ is not needed to
express $\ddot q$ in terms of the canonical variables. It appears
only linearly in the term $p_1q_2$ and therefore $H$ cannot be
bounded from below; i.e. the system is unstable. Of course it is
possible to find well behaved solutions of this system, since for
a given solution energy is conserved. But as soon as the system is
interacting, e.g. with a harmonic oscillator, it will lower its
energy and produce more and more oscillating modes.

This is especially serious when one quantizes the system. The
vacuum is exponentially unstable to simultaneous production of
modes of positive and negative energy. Of course one cannot simply
`cut away' the negative energy solutions without violating
unitarity. And even if the theory under consideration is only a
low energy effective theory, it should at least be `unitary at low
energy'.

It is clear that introducing even higher derivatives only worsens
the situation, since the degree of the Euler-Lagrange equation is
enhanced by 2 with each new degree of freedom. Hence if the
Lagrangian has degree $2+n$, there are $n+1$ pairs of canonical
variables needed to describe the Hamiltonian, and of these only
$n+2$ are needed to invert the Lagrangian. Hence $n$ momenta
appear only linearly in the terms $p_j \dot q_j(q_1,\cdots,
q_{n+1}, p_{n+1})$, and the Hamiltonian has $n$ unstable
directions.

In this argument, it does not at all matter whether the degrees of
freedom we are discussing are fundamental or only low energy
effective degrees of freedom. Even if we modify the Hamiltonian at
high energies, the instability, which is a {\em low energy
problem}, will not disappear. There are only two ways out of the
Ostrogradski instability: Firstly, if the necessary condition that
$L$ be non-degenerate is not satisfied. The second possibility is
via constraints. In a system with $m$ constraints, one can in
principle eliminate $m$ variables. Hence if a $2+n$ order system
has $n$ constraints one might be able to eliminate all the
unstable directions. In practice, this has to be studied on a case
by case basis.

An important example for the dark energy problem is modified
gravity Lagrangians of the form
 \be
L=\sqrt{-g}f(R, R_{\mu\nu}R^{\mu\nu}, C_{\mu\nu\al\beta}
C^{\mu\nu\al\beta}) ~.
 \ee
Here $R_{\mu\nu}$ is the Ricci tensor, $C_{\mu\nu\al\beta} $ is
the Weyl tensor and $f(x_1,x_2,x_3)$ is an arbitrary (at least
three times differentiable) function. Since the curvature tensors
contain second derivatives of the metric, the resulting equations
of motion will in general be fourth order and Ostrogradski's
theorem applies. The usual Hamiltonian formulation of general
relativity leads to six independent metric components $g_{ij}$
which all acquire higher derivative terms. There is actually only
one way out, which is the case $\dd_2f=\dd_3f=0$, i.e., $f$ may
only depend on $R$.\footnote{Another (trivial) possibility is the
  addition of a Gauss Bonnet term, $L_{GB}= \sqrt{-g}\left(
  R^2-4R_{\mu\nu}R^{\mu\nu}+
  R_{\mu\nu\sigma\rho}R^{\mu\nu\sigma\rho}\right)$,
  which in four dimensions contributes only a surface term and does
  not enter the equations of motion. However, such a term becomes
  interesting in scalar-tensor theories of gravity where one may
  consider a contribution of the form $\phi L_{GB}$ to the
  Lagrangian.}
The reason is that in the Riemann scalar $R$, only a single
component of the metric contains second derivatives. In this case,
the consequent new degree of freedom can be fixed completely by
the $g_{00}$ constraint, so that the only instability in $f(R)$
theories is the usual one associated with gravitational collapse
(see~\cite{Woody}).

Therefore, the only acceptable generalizations of the
Einstein-Hilbert action of general relativity are $f(R)$ theories,
reviewed in the contribution~\cite{CapFran}.

If  the Ostradgradski theorem does not apply, we have still no
guarantee that the theory has no ghosts or that the potential
energy is bounded from below (no `serious' tachyon). The
limitation from the Ostragradski theorem, but also the ghost and
tachyon problem, can be cast in the requirement that the theory
needs to have an energy functional which is bounded from below.
This condition can certainly not disappear in a consistent low
energy version of a fundamental theory which satisfies it.

Like ghosts, the Ostrogradski instability can in principle be
cured by adding a term $\propto (\Phi/m)^2(\nabla\varphi)^2$ to an
unstable mode $\varphi$, where $\Phi$ is a very heavy particle
with mass $M\gg m$, which has been neglected in the low energy
approximation of the theory. However, this means that the full low
energy theory actually must contain a term $(M/m)^2
(\nabla\varphi)^2$. Consequences worked out within the low energy
theory neglecting this term can in general not be trusted. Only a
detailed case by case analysis can then reveal which low energy
results still apply and which ones are modified by the coupling to the
massive field $\Phi$.

Furthermore, the high energy cut-off will be given by some mass
scale, i.e. some Lorentz invariant energy scale of the theory, and
therefore the effective low energy theory should also admit a
Lorentz invariant Lagrangian. Lorentz invariance is not a high
energy phenomenon which can simply be lost at low energies.

What about superluminal motion and causality? We do not want to
require certain properties of the $S$ matrix of the low energy
theory, since the latter may not have a meaningful perturbative
quantum theory; like the 4-Fermi theory, it may not be
renormalizable. Furthermore, one can argue that in cosmology we do
have a preferred frame, the cosmological frame, hence
Lorentz-invariance is broken and we can simply demand that all
superluminal modes of a field propagate forward in cosmic time.
Then no closed signal curves are possible.

But this last argument is very dangerous. Clearly, most solutions
of a Lagrangian theory do break several or most of the symmetries
of the Lagrangian spontaneously. But when applying a Lorentz
transformation to a solution, we produce a new solution that, from
the point of view of the Lagrangian, has the same right of
existence. If some modes of a field propagate with superluminal
speed, this means that their characteristics are spacelike. The
condition that the mode has to travel forward in time with respect
to a certain frame implies that one has to use the retarded
Green's function in this frame. Since spacelike distances have no
frame-independent chronology, for spacelike characteristics this
is a frame-dependent statement. Depending on the frame of
reference, a given mode can represent a normal propagating degree
of freedom, or it can satisfy an elliptic equation, a constraint.

Furthermore, to make sure that the mode propagates forward with
respect to one fixed reference frame, one would have to use
sometimes the retarded, sometimes the advanced and sometimes a
mixture of both functions, depending on the frame of reference. In
a cosmological setting this can be done in a consistent way, but
it is far from clear that such a prescription can be unambiguously
implemented  for generic low energy solutions. Indeed in
Ref.~\cite{adams} a solution is sketched that would not allow
this, so that closed signal curves are again possible.

Therefore, we feel that Lorentz invariant low energy effective
Lagrangians which allow for superluminal propagation of certain
modes, have to be rejected. Nevertheless, this case is not as
clear-cut and there are opposing opinions in the literature,
e.g.~\cite{muki}.

With the advent of the `landscape'~\cite{land}, physicists have
begun to consider anthropic arguments to justify their theory,
whenever it fits the data. Even though the existence of life on
earth is an experimental fact, we consider this argument weak,
nearly tantamount to giving up physics: `Things are like they are
since otherwise we would not be here'. We nevertheless find it
important to inquire also from a purely theoretical point of view,
whether really `anything goes' for effective theories. In the
following sections we shall come back to the basic requirements
which we have outlined in this section.

\section{GENERAL RELATIVISTIC APPROACHES}
\label{s:de}

The ``standard" general relativistic interpretation of dark energy
is based on the cosmological constant as vacuum energy:
 \be
G_{\mu\nu}= 8\pi G\left[ T_{\mu\nu}+T^{\text{vac}}_{\mu\nu}
\right],~~T^{\text{vac}}_{\mu\nu}=-{\Lambda_{\rm eff} \over 8\pi
G}\,g_{\mu\nu}\,,
 \ee
where the vacuum energy-momentum tensor is Lorentz invariant. This
approach faces the problem of accounting for the incredibly small
and highly fine-tuned value of the vacuum energy, as summarized in
Eq.~(\ref{v1}).

String theory provides a tantalising possibility in the form of
the ``landscape" of vacua~\cite{land}. There appears to be a vast
number of vacua admitted by string theory, with a broad range of
vacuum energies above and below zero. This is discussed in the
contribution by Bousso~\cite{rafi}. The idea is that our
observable region of the universe corresponds to a particular
small positive vacuum energy, whereas other regions with greatly
different vacuum energies will look entirely different. This
multitude of regions forms in some sense a ``multiverse". This is
an interesting idea, but it is highly speculative, and it is not
clear how much of it will survive the further development of
string theory and cosmology.

An alternative view of LCDM is the interpretation of $\Lambda$ as
a classical geometric constant (see, e.g.,
Ref.~\cite{Padmanabhan:2006cj}), on a par with Newton's constant
$G$. Thus the field equations are interpreted in the geometrical
way,
 \be
G_{\mu\nu}+\Lambda g_{\mu\nu}= 8\pi G T_{\mu\nu}\,.
 \ee
In this approach, the small and fine-tuned value of $\Lambda$ is
no more of a mystery than the host of other fine-tunings in the
constants of nature. For example, more than a 2\% change in the
strength of the strong interaction means that no atoms beyond
hydrogen can form, so that stars and galaxies would not emerge.
However, this classical approach to $\Lambda$ does not evade the
vacuum energy problem -- it simply shifts that problem to ``why
does the vacuum not gravitate?" The idea is that particle physics
and quantum gravity will somehow discover a cancellation or
symmetry mechanism to explain why $\rho_{\text{vac}}=0$. This
would be a simpler solution than that indicated by the string
landscape approach, and would evade the disturbing anthropic
aspects of that approach. Nevertheless, it is not evident, whether
this distinction between $\La$ and $\rho_{\text{vac}}$ is really a
physical statement, or a purely theoretical statement that cannot
be tested by any experiments.

Within general relativity, various alternatives to LCDM have been
investigated.

\subsection{{\bf Dynamical dark energy: quintessence}}

Here we replace the constant $\Lambda/8\pi G$ by the energy
density of a scalar field $\varphi$, with Lagrangian
 \be
L_\varphi = \frac{1}{2}g^{\mu\nu}\dd_\mu\varphi \dd_\nu\varphi +
V(\varphi)\,,
 \ee
so that in a cosmological setting,
 \be
&& \rho_\varphi={1\over2}\dot{\varphi}^2+V(\varphi)\,,~ \quad
p_\varphi={1\over2}\dot{\varphi}^2-V(\varphi)\,,\\
&& \ddot\varphi+3H\dot\varphi+V'(\varphi)=0\,,\\
&& H^2 +\frac{K}{a^2} =\frac{8\pi G}{3}\left(\rho_r +\rho_m
+\rho_\varphi\right) \,.
 \ee
The field rolls down its potential and the dark energy density
varies through the history of the universe. ``Tracker" potentials
have been found for which the field energy density follows that of
the dominant matter component. This offers the possibility of
solving or alleviating the fine tuning problem of the resulting
cosmological constant. Although these models are insensitive to
initial conditions, they do require a strong fine-tuning of the
{\em parameters of the Lagrangian} to secure recent dominance of
the field, and hence do not evade the coincidence problem. More
generally, the quintessence potential, somewhat like the inflaton
potential, remains arbitrary, until and unless fundamental physics
selects a potential. There is currently no natural choice of
potential.

In conclusion, there is no compelling reason as yet to choose
quintessence above the LCDM model of dark energy. Quintessence
models do not seem more natural, better motivated or less
contrived than LCDM.  Nevertheless, they are a viable possibility
and computations are straightforward. Therefore, they remain an
interesting target for observations to shoot at. More details and
references can be found in the contribution~\cite{linder}.

\subsection{{\bf Dynamical dark energy: more general models}}

It is possiblew to  couple quintessence  to cold dark matter,
so that the energy conservation equations become
 \be
\dot\varphi\left[ \ddot\varphi+3H\dot\varphi+V'(\varphi)\right]
&=& J\,,\\
\dot{\rho}_m+3H\rho_m &=& -J\,,
 \ee
where $J$ is the energy exchange~\cite{Amen,Guo}.

Another possibility is a scalar field with non-standard kinetic term in
the Lagrangian, for example,
\begin{equation}\label{ns}
{L}_\varphi= F(\varphi,X)+V(\varphi)~\mbox{where}~ X\equiv {1\over
2}g^{\mu\nu}\partial_\mu\varphi \partial_\nu\varphi\,.
\end{equation}
The standard Lagrangian has $F(\varphi,X)=X$. Some of the
non-standard $F$ models may be ruled out on theoretical grounds.
An example is provided by ``phantom" fields, with negative kinetic
energy density (ghosts), $F(\varphi,X)=-X$. They have $w<-1$, so
that their energy density {\em grows} with expansion. This bizarre
behaviour is reflected in the instability of the quantum vacuum
for phantom fields.

Another example is ``k-essence" fields~\cite{kess}, which have
$F(\varphi,X) = \varphi^{-2}f(X)$. These theories have no ghosts,
and they can produce late-time acceleration. The sound speed of
the field fluctuations for the Lagrangian in Eq.~(\ref{ns}) is
\begin{equation}
{c_s^2}={F_{,X} \over F_{,X}+2XF_{,XX}}\,.
\end{equation}
For a standard Lagrangian, $c_s^2=1$. But for the class of $F$
that produce accelerating k-essence models, it turns out that
there is always an epoch during which $c_s^2>1$, so that these
models may be ruled out according to our causality requirement.
They violate standard causality~\cite{caus}.

For models not ruled out on theoretical grounds, there is the same
general problem as with quintessence, i.e. that no model is better
motivated than LCDM, none is selected by fundamental physics and
any choice of model is more or less arbitrary. Quintessence then
appears to at least have the advantage of simplicity -- although
LCDM has the same advantage over quintessence.

When investigating generic dark energy models we always have to
keep in mind that since both dark energy and dark matter are only
detected gravitationally, we can only measure the total energy
momentum tensor of the dark component,
 \be
T_{\mu\nu}^{\rm dark}=T_{\mu\nu}^{\rm de} + T_{\mu\nu}^{\rm dm} \,
.
 \ee
Hence, if we have no information on the equation of state of dark
energy, there is a degeneracy between the dark energy equation of
state $w(t)$ and $\Om_{\rm dm}$. Without additional assumptions,
we cannot measure either of them~\cite{martin}. This degeneracy
becomes even worse if we allow for interactions between dark
matter and dark energy.

\subsection{{\bf Dark energy as a nonlinear effect from
structure}}

As structure forms and the matter density perturbation becomes
nonlinear, there are two questions that are posed: (1)~what is the
back-reaction effect of this nonlinear process on the background
cosmology?; (2)~how do we perform a covariant and gauge-invariant
averaging over the inhomogeneous universe to arrive at the correct
FRW background? The simplistic answers to these questions are:
(1)~the effect is negligible since it occurs on scales too small
to be cosmologically relevant; (2)~in light of this, the
background is independent of structure formation, i.e., it is the
same as in the linear regime. A quantitative analysis is needed to
fully resolve both issues. However, this is very complicated
because it involves the nonlinear features of general relativity
in an essential way.

There have been claims that these simplistic answers are wrong,
and that, on the contrary, the effects are large enough to
mimic an accelerating universe. This would indeed be a dramatic and
satisfying resolution of the coincidence problem, without the need
for any dark energy field. Of course, the problem of why the
vacuum does not gravitate would remain. This issue is discussed in
the contribution~\cite{thomas}.

However, these claims have been disputed, and it is fair to say
that there is as yet no convincing demonstration that acceleration
could emerge naturally from nonlinear effects of structure
formation; see Refs.~\cite{Kolb:2005me} for some claims and
counter-claims. We should however note the possibility that
backreaction/ averaging effects could be significant, even if they
do not lead to acceleration.

It might also be possible that the universe around us resembles
more a spherically symmetric but inhomogeneous solution of
Einstein's equation, a Tolman-Bondi-Lema\^\i tre universe, than a
Friedmann-Lema\^\i tre universe. In this case, what appears as
cosmic acceleration to us can be explained within simple matter
models which only contain dust. However, this would imply that we
are situated very close to the centre of a huge (nearly) spherical
structure. Apart from violating the Copernican principle, this
poses another fine tuning problem. This idea is discussed in the
contribution~\cite{kari}.

\section{THE MODIFIED GRAVITY APPROACH: DARK GRAVITY}
\label{s:dg}

Late-time acceleration from nonlinear effects of structure
formation is an attempt, within general relativity, to solve the
coincidence problem without a dark energy field. The modified
gravity approach shares the assumption that there is no dark
energy field, but generates the acceleration via ``dark gravity",
i.e. a weakening of gravity on the largest scales, due to a
modification of general relativity itself.

Could the late-time acceleration of the universe be a
gravitational effect? (Note that this would also not remove the
problem of explaining why the vacuum energy does not gravitate.) A
historical precedent is provided by attempts to explain the
anomalous precession of Mercury's perihelion by a ``dark planet",
named Vulcan. In the end, it was discovered that a modification to
Newtonian gravity was needed.

As we have argued in Section~2, a consistent modification of
general relativity requires a covariant formulation of the field
equations in the general case, i.e., including inhomogeneities and
anisotropies. It is not sufficient to propose ad hoc modifications
of the Friedman equation, of the form
 \be
f(H^2) = {8\pi G\over 3} \rho ~~\mbox{or}~ ~ H^2  = {8\pi G\over
3} g(\rho) \,,
 \ee
for some functions $f$ or $g$. Apart from the fundamental problems
outlined in Section~2, such a relation allows us to compute the
supernova distance/ redshift relation using this equation -- but
we {\em cannot} compute the density perturbations without knowing
the covariant parent theory that leads to such a modified Friedman
equation. And we also cannot compute the solar system predictions.

It is very difficult to produce infrared corrections to general
relativity that meet all the minimum requirements:
\begin{itemize}

\item
Theoretical consistency in the sense discussed in Section~2.

\item
Late-time acceleration consistent with supernova data.

\item
A matter-dominated era with an evolution of the scale factor $a$
that is consistent with the requirements of structure formation.

\item
Density perturbations that are consistent with the observed matter
power spectrum, CMB anisotropies and weak lensing power spectrum.

\item
Stable static spherical solutions for stars and vacuum, and
consistency with terrestrial and solar system observational
constraints.

\item
Consistency with binary pulsar period data.

\end{itemize}

\subsection{{\bf Scalar-tensor theories}}

General relativity has a unique status as a theory where gravity
is mediated by a massless spin-2 particle, and the field equations
are second order. If we introduce modifications to the
Einstein-Hilbert action of the general form
 \be
\int d^4x\,\sqrt{-g}\,R ~\to~ \int d^4x\,\sqrt{-g}\,f(R,
R_{\mu\nu}R^{\mu\nu},C_{\mu\nu\alpha\beta}C^{\mu\nu\alpha\beta})\,,
 \ee
then the field equations become fourth-order, and gravity is
carried also by massless spin-0 and spin-1 fields. In order to
avoid the Ostrogradski instability discussed in Section~2, we
impose $f=f(R)$, and we assume $f''(R)\neq 0$. However, it turns
out to be extremely difficult for this simplified class of
modified theories to pass the observational and theoretical tests.
An example is~\cite{Capozziello:2003tk}
 \be
f(R)=R-{\mu \over R}\,.
 \ee
For $|\mu|\sim H_0^{4}$, this model achieves late-time
acceleration as the $\mu/R$ term starts to dominate. But the
model suffers from  nonlinear matter instabilities and
violation of solar system constraints~\cite{fr}.

Variations of $f(R)$ theories have been introduced to evade these
problems~\cite{newfr}. These are based on a ``chameleon" mechanism
to alter the modification of general relativity across the
boundary between a massive body and its vacuum exterior. Although
such mechanisms may be successful, the models look increasingly
unnatural and contrived -- and suffer from very strong
fine-tuning.

All $f(R)$ theories lead to just one fourth order
equation~\cite{Woody}. The corresponding additional degree of
freedom can be interpreted as a scalar field and in this sense,
$f(R)$ theories are mathematically equivalent to scalar-tensor
theories via
 \be
\psi &\equiv& f'(R)\,,~~
U(\psi) \equiv  \psi -f(R(\psi)) \,,\\
L &=& \frac{1}{16\pi G} \sqrt{-g}\left[\psi R +U(\psi)\right].
 \ee
This Lagrangian can be conformally transformed into ordinary
gravity with a scalar field, i.e. a quintessence model, via the
transformation
 \be  \label{e:defphi}
\tilde g_{\mu\nu} = \psi g_{\mu\nu}   \,,~~ \varphi = \sqrt{3\over
4\pi G}\ln\psi ~.
 \ee
In terms of $\tilde g_{\mu\nu}$ and $\varphi$ the Lagrangian then
becomes a standard scalar field Lagrangian,
 \be
\sqrt{-g}f(R) = \sqrt{-\tilde g}\left[\tilde R
+\frac{1}{2}\tilde g^{\mu\nu}\dd_\mu\varphi \dd_\nu\varphi
+V(\varphi) \right]\,,
 \ee
where
 \be
V(\varphi)= {1\over 16\pi G}\,{U(\psi(\varphi)) \over
\psi(\varphi)^2} \,.
 \ee
This example shows that modifying gravity (dark gravity) or
modifying the energy momentum tensor (dark energy) can be seen as
a different description of the same physics. Only the coupling of
the scalar field $\varphi$ to ordinary matter, shows that this
theory originates from a scalar-tensor theory of gravity -- and
this non-standard coupling reflects the fact that gravity is also
mediated by a spin-0 degree of freedom, in contrast to general
relativity with a standard scalar field.

More general scalar-tensor theories~\cite{Boisseau:2000pr}, which
may be motivated via low-energy string theory, have an action of
the form
 \be
\int d^4x\,\sqrt{-g}\left[ F(\psi)R+{1\over 2}
g^{\mu\nu}\partial_\mu\psi\partial_\nu\psi + U(\psi) \right],
 \ee
where $\psi$ is the spin-0 field supplementing the spin-2
graviton. In the context of late-time acceleration, these models
are also known as ``extended quintessence". Scalar-tensor theories
contain two functions, $F$ and $U$. This additional freedom allows
for greater flexibility in meeting the observational and
theoretical constraints. However, the price we pay is additional
complexity -- and arbitrariness. The $f(R)$ theories have one
arbitrary function, and here there are two, $F(\psi)$ and
$U(\psi)$. There is no preferred choice of these functions from
fundamental theory.

In summary, modifications of the Einstein-Hilbert action, which
lead to fourth-order field equations, either fail to meet  the minimum
requirements in the simplest cases, or contain more complexity and
arbitrary choices than quintessence models in general relativity.
Therefore, none of these models appears to be a serious competitor
to quintessence in general relativity.

\subsection{{\bf Brane-world models}}

We turn now to a class of brane-world models whose background is
no more complicated than that of LCDM, offering the promise of a
serious dark gravity contender. However, there are hidden
complexities and problems, as we will explain below.

\begin{figure}[!bth]
\begin{center}
\includegraphics[height=3in,width=4in]{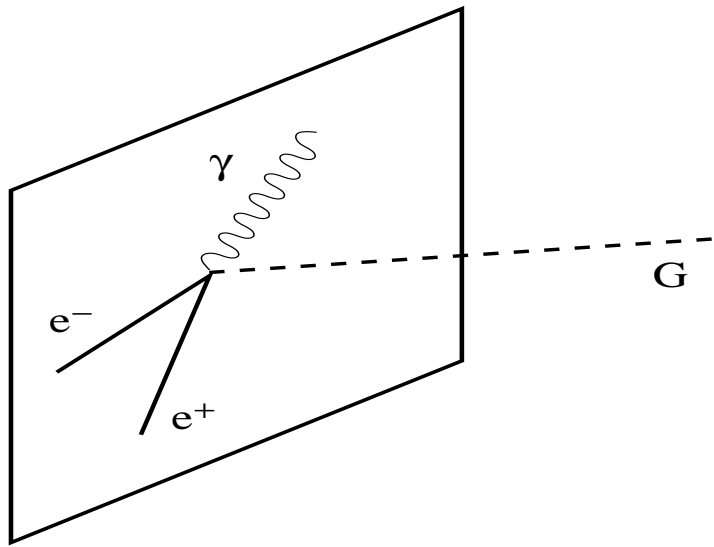}
\caption{The confinement of matter to the brane, while gravity
propagates in the bulk (from~\cite{Cavaglia:2002si}).}
\label{brane}
\end{center}
\end{figure}

An infra-red modification to general relativity can emerge within
the framework of quantum gravity, in addition to the ultraviolet
modification that must arise at high energies in the very early
universe. The leading candidate for a quantum gravity theory,
string theory, is able to remove the infinities of quantum field
theory and unify the fundamental interactions, including gravity.
But there is a price -- the theory is only consistent in 9 space
dimensions. Branes are extended objects of higher dimension than
strings, and play a fundamental role in the theory, especially
D-branes, on which open strings can end. Roughly speaking, the
endpoints of open strings, which describe the standard model
particles like fermions and gauge bosons, are attached to branes,
while the closed strings of the gravitational sector can move
freely in the higher-dimensional ``bulk" spacetime. Classically,
this is realised via the localization of matter and radiation
fields on the brane, with gravity propagating in the bulk (see
Fig.~\ref{brane}).

The implementation of string theory in cosmology is extremely
difficult, given the complexity of the theory. This motivates the
development of phenomenology, as an intermediary between
observations and fundamental theory. (Indeed, the development of
inflationary cosmology has been a very valuable exercise in
phenomenology.) Brane-world cosmological models inherit key
aspects of string theory, but do not attempt to impose the full
machinery of the theory. Instead, simplifications are introduced
in order to be able to construct cosmological models that can be
used to compute observational predictions
(see~\cite{Maartens:2003tw} for reviews in this spirit).
Cosmological data can then be used to constrain the brane-world
models, and hopefully provide constraints on string theory, as
well as pointers for the further development of string theory.

It turns out that even the simplest brane-world models are
remarkably rich -- and the computation of their cosmological
perturbations is complicated, and still incomplete. A key reason
for this is that the higher-dimensional graviton produces a tower
of 4-dimensional massive spin-2 modes on the brane, in addition to
the standard massless spin-2 mode on the brane (or in some cases,
instead of the massless mode). In the case of some brane models,
there are in addition a massless gravi-scalar and gravi-vector
which modify the dynamics.

Most brane-world models modify general relativity at high
energies. The main examples are those of Randall-Sundrum (RS)
type~\cite{Randall:1999vf}, where a FRW brane is embedded in an
anti de Sitter bulk, with curvature radius $\ell$. At low energies
$H\ell \ll 1$, the zero-mode of the graviton dominates on the
brane, and general relativity is recovered to a good
approximation. At high energies, small scales, $H\ell\gg 1$, the
massive modes of the graviton dominate over the zero mode, and
gravity on the brane behaves increasingly five dimensional. On the
brane, the standard conservation equation holds,
 \be
\dot\rho+3H(\rho+p)=0\,,
  \ee
but the Friedmann equation is modified by an ultraviolet
correction:
 \be\label{mf}
H^2 = \frac{8\pi G}{3} \rho\left(1+{2\pi G \ell^2\over 3}\rho
\right) + \frac{\Lambda}{3} \,.
 \ee
The $\rho^2$ term is the ultraviolet correction. At low energies,
this term is negligible, and we recover $H^2 \propto
\rho+\Lambda/8\pi G $. At high energies, gravity ``leaks" off the
brane and $H^2\propto \rho^2$. This 5D behaviour means that a
given energy density produces a greater rate of expansion than it
would in general relativity. As a consequence, inflation in the
early universe is modified in interesting
ways~\cite{Maartens:2003tw}.

By contrast, the brane-world model of
Dvali-Gabadadze-Porrati~\cite{Dvali:2000rv} (DGP), which was
generalized to cosmology by Deffayet~\cite{Deffayet:2000uy},
modifies general relativity at late times. This model produces
`self-acceleration' of the low-energy universe due to a weakening
of gravity. Like the RS model, the DGP model is a 5-dimensional
model with infinite extra dimension. (We effectively assume that 5
of the extra dimensions in the ``parent" string theory may be
ignored at low energies.)

The action is given by
 \be \label{DGPaction}
{1\over 16\pi G}\left[ {1\over r_c}\int_{\rm bulk}
d^5x\,\sqrt{-g^{(5)}}\,R^{(5)}+\int_{\rm brane} d^4x\,\sqrt{-g}\,R
\right] \,.
  \ee
The bulk is assumed to be 5D Minkowski spacetime. Unlike the AdS
bulk of the RS model, the Minkowski bulk has infinite volume.
Consequently, there is no normalizable zero-mode of the (bulk)
graviton in the DGP brane-world. Gravity leaks off the 4D brane
into the bulk at large scales, $\lambda>r_c$, where the first term
in the sum~(\ref{DGPaction}) dominates. On small scales, gravity
is effectively bound to the brane and 4D dynamics is recovered to
a good approximation, as the second term dominates. The transition
from 4- to 5D behaviour is governed by the crossover scale $r_c$;
the weak-field gravitational potential behaves as
\begin{equation}
\Psi \propto \left\{ \begin{array}{lll} r^{-1} & \mbox{for} & r\ll
r_c
\\ r^{-2} & \mbox{for} & r\gg r_c \end{array}\right.
\end{equation}
Gravity leakage at late times initiates acceleration -- not due to
any negative pressure field, but due to the weakening of gravity
on the brane. 4D gravity is recovered at high energy via the
lightest massive modes of the 5D graviton, effectively via an
ultra-light metastable graviton.

\begin{figure}[!bth]
\begin{center}
\includegraphics[height=3in,width=3in]{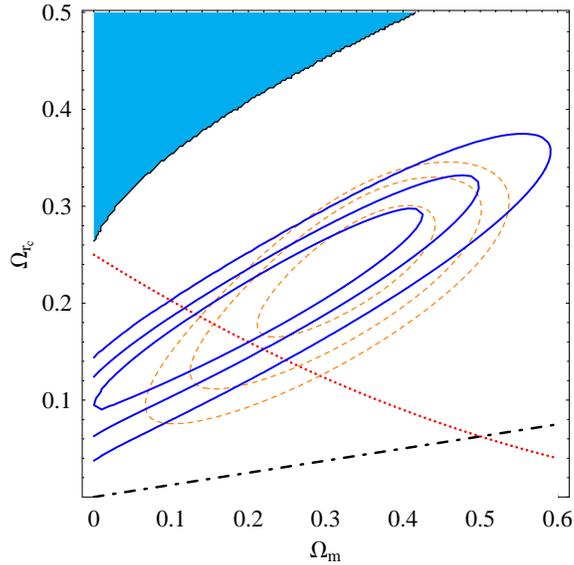}
\caption{The confidence contours for supernova data in the DGP
density parameter plane. The blue (solid) contours are for SNLS
data, and the brown (dashed) contours are for the Gold data. The
red (dotted) curve defines the flat models, the black (dot-dashed)
curve defines zero acceleration today, and the shaded region
contains models without a big bang. (From~\cite{mm}).}
\label{plane}
\end{center}
\end{figure}

The energy conservation equation remains the same as in general
relativity, but the Friedman equation is modified:
\begin{eqnarray}
\dot\rho+3H(\rho+p)&=&0\,,\label{ec} \\  H^2-{H \over r_c}&=&
{8\pi G \over 3}\rho\,. \label{f}
\end{eqnarray}
This shows that at early times, i.e., $Hr_c \gg 1$, the general
relativistic Friedman equation is recovered. By contrast, at late
times in a CDM universe, with $\rho\propto a^{-3}\to0$, we have
\begin{equation}
H\to H_\infty= {1\over r_c}\,,
\end{equation}
so that expansion accelerates and is asymptotically de Sitter.
Since $H_0>H_\infty$, in order to achieve self-acceleration at
late times, we require
 \be
r_c\gtrsim H_0^{-1}\,,
 \ee
and this is confirmed by fitting supernova observations, as shown
in Fig.~\ref{plane}. The dimensionless cross-over parameter is
 \be
\Omega_{r_c}={1\over 4(H_0r_c)^2}\,,
 \ee
and the LCDM relation,
 \be
\Omega_m+\Omega_\Lambda+\Omega_K=1\,,
 \ee
is modified to
 \be
\Omega_m+ 2\sqrt{\Omega_{r_c}}\sqrt{1-\Omega_K}+\Omega_K=1\,.
 \ee

\begin{figure}[t]
\centerline{
\includegraphics[width=9cm]{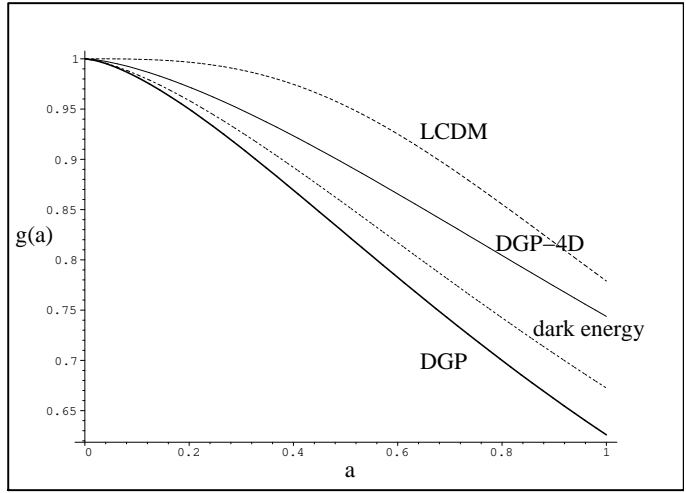}}
\caption{The growth factor $g(a)=\Delta(a)/a$ for LCDM (long
dashed) and DGP (solid, thick), as well as for a dark energy model
with the same expansion history as DGP (solid, thick). DGP-4D
(solid, thin) shows the incorrect result in which the 5D effects
are set to zero. (From~\cite{Koyama:2005kd}.)} \label{fig:fig1}
\end{figure}

LCDM and DGP can both account for the supernova observations, with
the fine-tuned values $\Lambda\sim H_0^2$ and $r_c\sim H_0^{-1}$
respectively. The degeneracy may be broken by observations based
on structure formation, since the two models suppress the growth
of density perturbations in different ways~\cite{Lue:2004rj}. The
distance-based observations draw only upon the background 4D
Friedman equation~(\ref{f}) in DGP models -- and therefore there
are quintessence models in general relativity that can produce
precisely the same supernova distances as
DGP~\cite{Linder:2005in}. By contrast, structure formation
observations require the 5D perturbations in DGP, and one cannot
find equivalent quintessence models~\cite{Koyama:2005kd}.
(However, 4D general relativity models allowing for anisotropic
stresses can in principle mimick DGP~\cite{KunzSap}.)

For LCDM, the analysis of density perturbations is well
understood. For DGP it is much more subtle and complicated. This
is discussed in the contribution~\cite{koyama}. Although matter is
confined to the 4D brane, gravity is fundamentally 5D, and the
bulk gravitational field responds to and back-reacts on density
perturbations. The evolution of density perturbations requires an
analysis based on the 5D nature of gravity. In particular, the 5D
gravitational field produces an anisotropic stress on the 4D
universe. If one neglects this stress and all 5D effects, and
simply treats the perturbations as 4D perturbations with a
modified background Hubble rate -- then as a consequence, the 4D
Bianchi identity on the brane is violated, i.e., $\nabla^\nu
G_{\mu\nu} \neq 0$, and the results are inconsistent.

When the 5D effects are incorporated~\cite{Koyama:2005kd,Card}, the 4D
Bianchi identity is satisfied. The consistent modified evolution
equation for density perturbations on sub-Hubble scales is
\begin{equation}\label{dpe}
\ddot{\Delta} + 2 H \dot{\Delta}=4\pi G \left\{1 -
\frac{(2Hr_c-1)}{3[2(Hr_c)^2-2 Hr_c+1]} \right\} \rho \Delta\,,
\end{equation}
where the term in braces encodes the 5D correction. The linear
growth factor, $g(a)=\Delta(a)/a$ (i.e., normalized to the flat
CDM case, $\Delta \propto a$), is shown in Fig.~\ref{fig:fig1}.

In addition to the complexity of the cosmological perturbations, a
deeper problem is posed by the fact that the late-time asymptotic
de Sitter solution in DGP cosmological models has a
ghost~\cite{Gorbunov:2005zk}. This ghost in the gravitational
sector is more serious than the ghost in a phantom scalar field.
It is actually this ghost degree of freedom which is responsible
for acceleration in the DGP model. Nevertheless, it may still
be usefull to study DGP as a toy model for dark gravity.

\section{CONCLUSION}
\label{s:con}

The evidence for a late-time acceleration of the universe
continues to mount, as the number of experiments and the quality
of data grow -- dark energy or dark gravity appear to be an
unavoidable reality of the cosmos. This revolutionary discovery by
observational cosmology, confronts theoretical cosmology with a
major crisis -- how to explain the origin of the acceleration. The
core of this problem may be ``handed over" to particle physics,
since we require at the most fundamental level, an explanation for
why the vacuum energy either has an incredibly small and
fine-tuned value, or is exactly zero. Both options violently
disagree with naive estimates of the vacuum energy.

If one accepts that the vacuum energy is indeed nonzero, then the
dark energy is described by $\Lambda$, and the LCDM model is the
best current model. The cosmological model requires completion via
developments in particle physics that will explain the value of
the vacuum energy. In many ways, this is the best that we can do
currently, since the alternatives to LCDM, within and beyond
general relativity, do not resolve the vacuum energy crisis, and
furthermore have no convincing theoretical motivation. None of the
contenders appears any better than LCDM.

Presently, perhaps the simplest and most appealing contender is
the DGP brane-world model. However, the simplicity of its Friedman
equation is deceptive, and the complexity of its cosmological
perturbations includes the problem of its ghost.

In view of all this, it is fair to say that at the theoretical
level, there is as yet no serious challenger to LCDM. It remains
worthwhile to continue investigating alternative dark energy and
dark gravity models, in order better to understand the space of
possibilities, the variety of cosmological properties, and the
observational strategies needed to distinguish them.

At the same time, it is in principle possible that cosmological
observations, having discovered dark energy/ dark gravity, could
rule out LCDM, by showing, to some acceptable level of statistical
confidence, that $w\neq -1$.

Finally, the theoretical crisis does not have only negative
implications: dark energy/ dark gravity in the cosmos provides
exciting challenges for theory and observations.

\begin{acknowledgements}
We thank Camille Bonvin, Chiara Caprini, Kazuya Koyama, Martin Kunz,
  Sanjeev Seahra and Norbert Straumann
  for stimulating and illuminating discussions. This work
  is supported by the Swiss National Science Foundation and the UK STFC.
\end{acknowledgements}

\end{document}